\documentclass[11pt]{article}
\usepackage{epsfig} 
\usepackage{amssymb}
\newcommand{\sect}[1]{ \section{#1} \setcounter{equation}{0} } 

\newcommand{\half}{\mbox{\small{$\frac{1}{2}$}}}

\newcommand{\MSbar}{\overline{\mbox{MS}}}

\begin{document}
\title{$F_4$ symmetric $\phi^3$ theory at four loops}
\author{J.A. Gracey, \\ Theoretical Physics Division, \\ 
Department of Mathematical Sciences, \\ University of Liverpool, \\ P.O. Box 
147, \\ Liverpool, \\ L69 3BX, \\ United Kingdom.} 
\date{}
\maketitle 

\vspace{5cm} 
\noindent 
{\bf Abstract.} The renormalization group functions for six dimensional scalar
$\phi^3$ theory with an $F_4$ symmetry are provided at four loops in the
modified minimal subtraction ($\MSbar$) scheme. Aside from the anomalous
dimension of $\phi$ and the $\beta$-function this includes the mass operator 
and a $\phi^2$-type operator. The anomalous dimension of the latter is computed
explicitly at four loops for the {\bf 26} and {\bf 324} representations of 
$F_4$. The $\epsilon$ expansion of all the related critical exponents are 
determined to $O(\epsilon^4)$. For instance the value for $\Delta_\phi$ agrees 
with recent conformal bootstrap estimates in $5$ and $5.95$ dimensions. The 
renormalization group functions are also provided at four loops for the group 
$E_6$. 

\vspace{-17cm}
\hspace{13cm}
{\bf LTH 1124}

\newpage 

\sect{Introduction.}

The mid 1980's saw a revolution in our understanding of two dimensional field
theories due to the development and classification of conformal field theories,
\cite{1}. The extension beyond strictly two dimensions has not been as 
straightforward mainly due to the different structure of the underlying 
conformal group in two dimensions and $d$~$>$~$2$ where $d$ is the spacetime
dimension. One recent development which is very promising is the so-called
conformal bootstrap programme, \cite{2,3,4,5}, which extended original ideas
on higher dimensional conformal theories, \cite{6,7,8,9,10,11,12}. Based on the
earlier work of \cite{2}, the conformal bootstrap has led to a new way of
estimating critical exponents in field theories in $d$~$>$~$2$. One primary 
example of the bootstrap success is in the three dimensional Ising model, 
\cite{4}, where estimates of exponents are competitive with other approaches 
such as strong coupling expansions, high temperature expansion and the 
$\epsilon$-expansion derived from perturbative renormalization group functions.
A comprehensive review and summary of the results from these methods is given
in \cite{13}. A topic which has subsequently been part of this development is 
the study of scalar field theories at the Wilson-Fisher fixed point in 
dimensions greater than four. In \cite{14,15} it was demonstrated that $O(N)$ 
$\phi^4$ theory could be extended into the $4$~$<$~$d$~$<$~$6$ range of 
spacetime dimensions and was in the same universality class as the $O(N)$ 
scalar $\phi^3$ theory which is perturbatively renormalizable in six 
dimensions. As well as the application of exact and functional renormalization 
group methods to this $O(N)$ theory conformal bootstrap studies also ensued 
with the main focus being the five dimensional theory, 
\cite{16,17,18,19,20,21,22}. For instance, there has been a debate as to where 
the boundary of the conformal window, akin to that determined by the Banks-Zaks
fixed point in gauge theories, \cite{23}, actually is. Bootstrap and functional
renormalization group methods have yet to arrive at a consensus for even the 
ballpark area for the conformal window boundary $N_{\mbox{\footnotesize{cr}}}$,
\cite{16,17,18,19,20,21,22}. Some bootstrap approaches find a low value for 
$N_{\mbox{\footnotesize{cr}}}$ but others suggest a value in keeping with the 
estimates from the $\epsilon$ expansion at four loops, \cite{14,15,24}, which 
is in the neighbourhood of $N$~$=$~$400$. For instance, this value is not 
inconsistent with the mixed correlator bootstrap estimate of \cite{22}. While 
such a discrepancy between different methods for $N_{\mbox{\footnotesize{cr}}}$
has yet to be resolved what is not in question is that critical exponent 
estimates are in broad agreement. This is reassuring as ultimately if all 
methods had access to tools to refine their computations then they ought to 
agree precisely.   
 
While the bootstrap debate to a degree has centred on quantum field theories
with a classical Lie group symmetry, a recent study has concentrated on the
exceptional group $F_4$, \cite{25}, as well as a brief look at the case of
$E_6$ symmetry. This was partly to complement the study of the 
$d$~$=$~$6$~$-$~$2\epsilon$ dimensional infrared stable fixed point with an
$O(N)$ symmetry which can access the five dimensional theory. Clearly the issue
of a conformal window is absent in the $F_4$ context in the sense that there is
no range of a group parameter for which there is a Banks-Zaks type fixed point.
However, \cite{25} also provided another forum to explore the conformal 
bootstrap technology. Indeed in \cite{25} the renormalization group functions 
for the six dimensional cubic $F_4$ symmetric field theory were determined to 
three loops in the modified minimal subtraction ($\MSbar$) scheme. These were 
derived from the earlier results of \cite{26,27,24}. One interesting outcome of
\cite{25} was the estimate for the field anomalous dimension in $d$~$=$~$5.95$ 
dimensions which was in precise agreement with that of the three loop 
$\epsilon$ expansion computed in perturbation theory. The study in $d$~$=$~$5$ 
dimensions was less clear in that the three loop estimate appeared to be out of 
line with that from the conformal bootstrap. While it was suggested this was 
due to non-perturbative effects, \cite{25}, one way to clarify this would be to
extend the three loop $F_4$ perturbative renormalization group functions to 
four loops. This is the purpose of this article. We will determine the 
$\beta$-function and field and mass anomalous dimensions to four loops in the 
$\MSbar$ scheme. However, as the bootstrap study in \cite{25} involved other 
$\phi^2$-type operators we will compute their anomalous dimensions to four 
loops as well. Their one loop terms were given in \cite{25}. These extra 
operators are variants of the mass operator but in the {\bf 26} and {\bf 324} 
representations of $F_4$. They are required in the application of the operator 
product expansion decomposition of the product of fields into conformal primary
operators in order to set up equations which the bootstrap technology solves. 
It turns out that having the critical exponents associated with these operators
to the same level of accuracy gives insight into the interpretation of the 
estimate for the $d$~$=$~$5$ dimensional exponent for the field as well as the 
$\phi^2$-type operators. Indeed there is a suggestion that perturbative results
from the underlying renormalization group functions could be used in future 
bootstrap studies as an aid or guide. For instance, in the $F_4$ case the 
exponent estimates derived from the $\epsilon$ expansion determine the order 
the operators appear in the spectrum in relation to increasing value. In 
addition we will provide the same renormalization group functions to four loops
for other Lie groups in the family with underlying $F_4$ symmetry as well as 
the exceptional group $E_6$.

The article is organized as follows. We briefly recap the key aspects of the
cubic scalar field theory in six dimensions upon which are computations are
based in section $2$. This includes the definition of the $\phi^2$-type 
operators and an outline of the method we used to efficiently determine their
four loop anomalous dimensions. To achieve this we need to use properties of 
the $F_4$ Lie algebra in order to evaluate the group factors associated with
each Feynman graph. This is discussed in section $3$ prior to the presentation
of the renormalization group functions in section $4$. Estimates for the 
critical exponents are also given there for the two specific dimensions of
interest. Section $5$ is devoted to the same analysis for the group $E_6$ while
we provide conclusions in section $6$. An appendix records the various critical
exponents for the family of groups with related $F_4$ symmetry.

\sect{Background.}

We begin by briefly recalling the necessary properties of the cubic scalar
field theory in six dimensions. Our massless Lagrangian is 
\begin{equation}
L ~=~ \frac{1}{2} \left( \partial_\mu \phi^i \right)^2 ~+~
\frac{g}{6} d^{ijk} \phi^i \phi^j \phi^k
\label{lag}
\end{equation}
where $d^{ijk}$ is totally symmetric and the group indices have the range
$1$~$\leq$~$i$~$\leq$~$N$. We use the same coupling constant conventions as 
\cite{24}. (As an aside we note the work of \cite{28} where the exponent $\eta$
of the single field $\phi^3$ theory was computed directly using conformal 
bootstrap methods of \cite{29} and can be regarded as a check on the results of
\cite{24}.) The mass operator 
\begin{equation}
{\cal O} ~=~ \frac{1}{2} \phi^i \phi^i
\end{equation}
has been omitted as the new aspect of the renormalization of (\ref{lag}) here 
is that we will consider a set of $\phi^2$-type operators in different group 
representations and compute their anomalous dimensions to four loops. The 
anomalous dimension of one element of this set of operators is equivalent to 
that of ${\cal O}$ which is already known to four loops. To be more specific in
order to build the various operators in a representation of $F_4$ there are 
five independent combinations of products of the group tensors $\delta^{ij}$ 
and $d^{ijk}$ with four free indices. These are $\delta^{ij} \delta^{kl}$, 
$\delta^{ik} \delta^{jl}$, $\delta^{il} \delta^{jk}$, $d^{ikp} d^{jlp}$ and 
$d^{ilp} d^{jkp}$. The additional product of $d^{ijp} d^{klp}$ is not 
independent due to the $F_4$ $4$-term relation. The various linear combinations
of these tensors which correspond to projectors onto the $F_4$ representations 
were given in \cite{30}. These lead to the set of rank $2$ $\phi^2$-type 
operators  
\begin{equation}
{\cal O}_{ij}^{(R)} ~=~ \frac{1}{2} {\cal P}^{(R)}_{ijkl} \phi^k \phi^l
\end{equation}
where
\begin{eqnarray}
{\cal P}^{(1)}_{ijkl} &=& \frac{1}{N} \delta_{ij} \delta_{kl} ~~~,~~~
{\cal P}^{(26)}_{ijkl} ~=~ d_{ijp} d_{klp} \nonumber \\
{\cal P}^{(324)}_{ijkl} &=& \frac{T_2}{2} \left[ \delta_{ik} \delta_{jl} ~+~ 
\delta_{il} \delta_{jk} ~-~ \frac{2}{N} \delta_{ij} \delta_{kl} \right] ~-~
d_{ijp} d_{klp} \nonumber \\
{\cal P}^{(52)}_{ijkl} &=& \frac{8}{[N+10]} \left[ \frac{T_2}{2} \left[ 
\delta_{il} \delta_{jk} ~-~ \delta_{ik} \delta_{jl} \right] ~+~ 
\frac{[N+2]}{8} \left[ d_{ilp} d_{jkp} ~-~ d_{ikp} d_{jlp} \right] \right]
\nonumber \\
{\cal P}^{(273)}_{ijkl} &=& \frac{[N+2]}{[N+10]} \left[ \frac{T_2}{2} \left[ 
\delta_{il} \delta_{jk} ~-~ \delta_{ik} \delta_{jl} \right] ~-~ 
d_{ilp} d_{jkp} ~+~ d_{ikp} d_{jlp} \right]
\end{eqnarray}
are the projectors in $F_4$, \cite{30,25}. Consequently the anomalous dimension
of ${\cal O}$ and ${\cal O}^{(1)}_{ij}$ are equivalent. Also the anomalous 
dimensions for the {\bf 52} and {\bf 273} representations are immediately zero 
since the operator $\half \phi^k \phi^l$ is symmetric whereas the respective 
projectors are antisymmetric in the indices $k$ and $l$. So our focus here will
be on establishing the four loop anomalous dimensions of the symmetric {\bf 26}
and {\bf 324} representations of $F_4$.

To determine the four loop anomalous dimensions of these operators we follow
the same method as outlined in \cite{24}. First, to renormalize an operator 
which does not mix it is inserted into the Green's function $\langle \phi^i(p) 
{\cal O}^{(R)}_{kl}(q) \phi^j(-p-q) \rangle$ where the external momenta $p$ and
$q$ flow in through the external legs. Then evaluating the constituent Feynman 
graphs the renormalization constant for the operator is deduced from the poles
in the regulator. In this paper we have used dimensional regularization in
$d$~$=$~$6$~$-$~$2\epsilon$ where $\epsilon$ is the regularizing parameter. 
Moreover we will use the modified minimal subtraction ($\MSbar$) scheme to
define the renormalization constants. While this outlines the essence of the
standard operator renormalization procedure for the renormalization of the
$\phi^2$-type operators defined from (\ref{lag}) there are several technical
shortcuts which allow us to extract the {\em four} loop renormalization
constants. As the operators do not involve derivatives then the insertion in
the Green's function can be at zero momentum. If one was considering the 
renormalization of the mass operator in four dimensions then this nullification
would be problematic. This is because the Feynman integrals would contain 
$1/(k^2)^2$, where $k$ is a loop momentum, which is infrared divergent. In six
dimensions, however, such a double pole propagator in an integral is infrared 
safe. So inserting the $\phi^2$-type operators at zero momentum will not 
corrupt the emergent operator renormalization constant with infrared 
divergences. Therefore we have relegated the exercise of renormalizing 
${\cal O}_{ij}^{(R)}$ in effect to one of evaluating a $2$-point function. As 
noted in \cite{24} this could involve $540$ Feynman diagrams to determine, for 
instance, for the nullified $3$-point function. However, for the 
renormalization of ${\cal O}$ the Green's function $\langle \phi^i(p) 
{\cal O}(0) \phi^j(-p) \rangle$ was generated from the graphs of the 
$\phi^i$ $2$-point function and we followed the same process here. For each 
graph of the $\phi^i$ $2$-point function one applies the map 
\begin{equation}
\frac{\delta_{ij}}{k^2} ~\mapsto~ \frac{\delta_{ij}}{k^2} ~+~
\frac{m^2}{2(k^2)^2} \left[ {\cal P}^{(R)}_{ijk_e l_e} ~+~ 
{\cal P}^{(R)}_{ijl_e k_e} \right] 
\end{equation}
to each propagator where we have not made any assumptions on the symmetry
properties of the projection tensor. The particular combination which appears
derives from the Feynman rule for ${\cal O}^{(R)}_{ij}$. The quantity $m$ is 
not a mass as such but a counting parameter. After the substitution has been 
made one truncates the graphs by retaining terms up to and including $O(m^2)$ 
only. Terms higher in $m^2$ would correspond to more than one insertion of the 
operator and correspond to a Green's function we are not interested in. The 
indices $k_e$ and $l_e$ are those associated with the external indices of the 
operator insertion. The advantage of using this technique to generate the 
particular Green's function is that it is straightforward to implement within 
our automatic Feynman diagram calculation. The graphs for the $\phi^i$ 
$2$-point function are generated with the package {\sc Qgraf}, \cite{31}, and 
converted into the syntax of the symbolic manipulation language we use which is 
{\sc Form} and its multithreaded version {\sc Tform}, \cite{32,33}. As the 
computation we performed to determine the anomalous dimensions of the {\bf 26} 
and {\bf 324} representations of ${\cal O}^{(R)}$ used the same programmes as 
that for ${\cal O}$ we refer the interested reader to \cite{24} for the 
technical details where the use of the Laporta algorithm, \cite{34}, and its
implementation in {\sc Reduze}, \cite{35,36}, is discussed in depth. The only 
major difference is that we have had to develop a {\sc Form} module to handle 
the group theory associated with the $F_4$ tensor $d^{ijk}$. 

\sect{$F_4$ group theory.}

We devote this section to the mechanics of finding the values for the group
invariants which appear in the $\phi^3$ theory renormalization to four loops
inclusive. These were defined in \cite{26,27} to three loops and the four loop
ones were introduced in \cite{24}. We use the notation introduced in the latter
and for completeness we note that they are
\begin{eqnarray}
&& d^{i i_1 i_2} d^{j i_1 i_2} ~=~ T_2 \delta^{i j} ~~,~~ 
d^{i i_1 i_2} d^{j i_1 i_3} d^{k i_2 i_3} ~=~ T_3 d^{i j k} ~~,~~ 
d^{i i_1 i_2} d^{j i_3 i_4} d^{k i_5 i_6} d^{i_1 i_3 i_5} 
d^{i_2 i_4 i_6} ~=~ T_5 d^{i j k} \nonumber \\
&& d^{i i_1 i_2} d^{j i_3 i_4} d^{k i_5 i_6} d^{i_1 i_3 i_7} d^{i_2 i_5 i_8} 
d^{i_4 i_6 i_9} d^{i_7 i_8 i_9} ~=~ T_{71} d^{i j k} \nonumber \\
&& d^{i i_1 i_2} d^{j i_3 i_4} d^{k i_5 i_6} d^{i_1 i_3 i_7} d^{i_2 i_5 i_8} 
d^{i_4 i_8 i_9} d^{i_6 i_7 i_9} ~=~ T_{72} d^{i j k} \nonumber \\
&& d^{i i_1 i_2} d^{j i_3 i_4} d^{k i_5 i_{12}} d^{i_1 i_5 i_6}
d^{i_2 i_7 i_8} d^{i_3 i_9 i_{12}} d^{i_4 i_7 i_{10}}
d^{i_6 i_8 i_{11}} d^{i_9 i_{10} i_{11}} ~=~ T_{91} d^{i j k} \nonumber \\
&& d^{i i_1 i_2} d^{j i_3 i_4} d^{k i_{11} i_{12}}
d^{i_1 i_5 i_6} d^{i_2 i_7 i_8} d^{i_3 i_5 i_9}
d^{i_4 i_7 i_{10}} d^{i_6 i_8 i_{11}} d^{i_9 i_{10} i_{12}} ~=~ T_{92} 
d^{i j k} \nonumber \\
&& d^{i i_1 i_2} d^{j i_3 i_4} d^{k i_6 i_{12}}
d^{i_1 i_5 i_6} d^{i_2 i_7 i_8} d^{i_3 i_5 i_9}
d^{i_4 i_7 i_{10}} d^{i_8 i_{11} i_{12}} d^{i_9 i_{10} i_{11}} ~=~ 
T_{93} d^{i j k} \nonumber \\
&& d^{i i_1 i_2} d^{j i_3 i_4} d^{k i_5 i_{12}}
d^{i_1 i_5 i_6} d^{i_2 i_7 i_8} d^{i_3 i_9 i_{12}}
d^{i_4 i_{10} i_{11}} d^{i_6 i_7 i_{10}} d^{i_8 i_9 i_{11}} ~=~ 
T_{94} d^{i j k} \nonumber \\
&& d^{i i_1 i_2} d^{j i_3 i_4} d^{k i_8 i_{12}}
d^{i_1 i_5 i_6} d^{i_2 i_7 i_8} d^{i_3 i_5 i_9}
d^{i_4 i_{10} i_{11}} d^{i_6 i_7 i_{10}} d^{i_9 i_{11} i_{12}} ~=~ 
T_{95} d^{i j k} \nonumber \\
&& d^{i i_1 i_2} d^{j i_3 i_4} d^{k i_{11} i_{12}}
d^{i_1 i_3 i_5} d^{i_2 i_6 i_7} d^{i_4 i_6 i_8}
d^{i_5 i_9 i_{10}} d^{i_7 i_9 i_{11}} d^{i_8 i_{10} i_{12}} ~=~ 
T_{96} d^{i j k} \nonumber \\
&& d^{i i_1 i_2} d^{j i_3 i_4} d^{k i_5 i_{12}}
d^{i_1 i_5 i_6} d^{i_2 i_7 i_8} d^{i_3 i_9 i_{12}}
d^{i_4 i_7 i_{10}} d^{i_6 i_{10} i_{11}} d^{i_8 i_9 i_{11}} ~=~ 
T_{97} d^{i j k} \nonumber \\
&& d^{i i_1 i_2} d^{j i_3 i_4} d^{k i_6 i_{12}}
d^{i_1 i_5 i_6} d^{i_2 i_7 i_8} d^{i_3 i_5 i_9}
d^{i_4 i_7 i_{10}} d^{i_8 i_9 i_{11}} d^{i_{10} i_{11} i_{12}} ~=~ 
T_{98} d^{i j k} \nonumber \\
&& d^{i i_1 i_2} d^{j i_3 i_4} d^{k i_{11} i_{12}}
d^{i_1 i_5 i_6} d^{i_2 i_7 i_8} d^{i_3 i_5 i_9}
d^{i_4 i_7 i_{10}} d^{i_6 i_{10} i_{11}} d^{i_8 i_9 i_{12}} ~=~ 
T_{99} d^{i j k} ~.
\label{gpinvtdef}
\end{eqnarray}
In that article values were derived for certain groups and it transpired that 
several subsets of $T_{9i}$ had the same values. We do not assume at the outset
that the same feature arises for either $F_4$ or $E_6$ which is considered 
later.

As the first stage in the extraction of the four loop renormalization group
functions for the $F_4$ symmetric case we recall basic properties of the tensor
$d^{ijk}$ for this specific group. From \cite{25,30} the $4$-term relation is  
\begin{equation}
d^{i j i_1} d^{k l i_1} ~+~ d^{i k i_1} d^{j l i_1} ~+~ 
d^{i l i_1} d^{j k i_1} ~=~ \left[ \delta^{i j} \delta^{k l} ~+~ 
\delta^{i k} \delta^{j l} ~+~ \delta^{i l} \delta^{j k} \right] 
\frac{2 T_2}{[N+2]}
\label{4term}
\end{equation}
where we retain $T_2$ for completeness in contrast to \cite{25}. From this we 
can derive an expression for the group theory value of a one loop box graph 
which we use extensively throughout. Although the same relation was given in
\cite{30} our derivation differs slightly from that given in \cite{25,30} but 
may be useful for constructing parallel expressions for the group theory 
associated with higher point one loop graphs. Contracting (\ref{4term}) with 
$d^{jpr} d^{kqr}$ produces 
\begin{eqnarray}
d^{i i_1 i_2} d^{j i_3 i_2} d^{k i_1 i_4} d^{l i_3 i_4} ~+~ 
d^{i i_1 i_2} d^{j i_3 i_2} d^{k i_3 i_4} d^{l i_1 i_4} &=& 
\left[ d^{i k i_1} d^{j l i_1} ~+~ d^{i l i_1} d^{j k i_1} \right]
\frac{2T_2}{[N+2]} \nonumber \\
&& +~ \delta^{i j} \delta^{k l} \frac{2T_2^2}{[N+2]} \nonumber \\
&& +~ d^{i j i_1} d^{k l i_1} \frac{[N-2]T_2}{2[N+2]} 
\label{boxcomb}
\end{eqnarray}
after relabelling. We have used $d^{ijj}$~$=$~$0$ which follows from 
contracting (\ref{4term}) with $d^{ijp}$. The two terms on the left hand side 
of (\ref{boxcomb}) represent two permutations of the indices on a one loop box 
diagram. If we define the tensor
\begin{equation}
B^{ijkl} ~=~ d^{i i_1 i_2} d^{j i_3 i_2} d^{k i_1 i_4} d^{l i_3 i_4} 
\end{equation}
to denote the group theory associated with a one loop box integral and the 
right hand side of (\ref{boxcomb}) formally by $C^{ijkl}$ then (\ref{boxcomb}) 
becomes 
\begin{equation}
B^{ijkl} ~+~ B^{ijlk} ~=~ C^{ijkl} ~.
\label{box1}
\end{equation} 
By permuting the indices $j$~$\to$~$k$, $k$~$\to$~$l$ and $l$~$\to$~$j$ twice 
we obtain two other relations which are
\begin{equation}
B^{ijlk} ~+~ B^{ikjl} ~=~ C^{iklj} ~~~,~~~  
B^{ikjl} ~+~ B^{ijkl} ~=~ C^{iljk} 
\label{box2}
\end{equation} 
where we have used $B^{ijkl}$~$=$~$B^{ilkj}$ which follows from the definition 
of $B^{ijkl}$. These three relations, (\ref{box1}) and (\ref{box2}), involve 
the three independent ways of labelling a one loop box topology and hence the
equations can be represented by a $3$~$\times$~$3$ matrix where the entries are
either $0$ or $1$. Inverting this matrix then allows one to obtain the 
decomposition of a box topology into tensors involving fewer products of 
$d^{ijk}$ tensors which is, \cite{25,30},  
\begin{eqnarray}
d^{i i_1 i_3} d^{j i_2 i_3} d^{k i_2 i_4} d^{l i_1 i_4} &=& 
\left[ \delta^{i j} \delta^{k l} ~-~ \delta^{i k} \delta^{j l}
~+~ \delta^{i l} \delta^{j k} \right] \frac{T_2^2}{[N+2]} \nonumber \\
&& +~ \left[ d^{i j i_1} d^{k l i_1} ~+~ 
d^{i l i_1} d^{j k i_1} \right] \frac{[N-2] T_2}{4[N+2]} \nonumber \\
&& -~ d^{i k i_1} d^{j l i_1} \frac{[N-10] T_2}{4[N+2]} 
\label{boxdecomp}
\end{eqnarray}
where the right hand side is derived from $C^{ijkl}$ and its above 
permutations. The result is equivalent to that produced in \cite{30} using a 
symmetrization and antisymmetrization method. Equipped with this has allowed us
to evaluate all the group invariants of (\ref{gpinvtdef}) aside from $T_2$. Its
value will not be set to unity until later. We find  
\begin{eqnarray}
T_3 &=& -~ [ N - 2 ] \frac{T_2}{2[N+2]} ~~,~~ 
T_5 ~=~ -~ [ N^2 - 10 N - 16 ] \frac{T_2^2}{2[N+2]^2} \nonumber \\
T_{71} &=& [ N^3 - 3 N^2 + 80 N + 100 ] \frac{T_2^3}{4[N+2]^3} ~~,~~ 
T_{72} ~=~ -~ N [ N - 2 ] [ N - 10 ] \frac{T_2^3}{8[N+2]^3} \nonumber \\
T_{91} &=& [ N - 2 ] [ N^3 + 18 N^2 + 204 N + 152 ] \frac{T_2^4}{16[N+2]^4} 
\nonumber \\
T_{92} &=& T_{96} ~=~ T_{98} ~=~ -~ [ N - 2 ] [ N^3 - 24 N^2 + 36 N + 80 ] 
\frac{T_2^4}{8[N+2]^4} \nonumber \\
T_{93} &=& T_{95} ~=~ T_{97} ~=~ N [ N - 2 ] [ N^2 - 24 N + 12 ] 
\frac{T_2^4}{16[N+2]^4} \nonumber \\
T_{94} &=& -~ [ N^4 - 14 N^3 - 12 N^2 - 616 N - 672 ] \frac{T_2^4}{8[N+2]^4} 
\nonumber \\
T_{99} &=& [ N - 2 ] [ N^3 - 27 N^2 + 54 N + 72 ] \frac{T_2^4}{4[N+2]^4} ~.
\label{gpinvf4}
\end{eqnarray}
As a check we have reproduced the expressions for the invariants used in the 
three loop analysis of \cite{25}. Those at four loops, $T_{9n}$, are new. The
only one where we could not directly reduce the invariant using 
(\ref{boxdecomp}) was $T_{99}$. This was because all the one loop subgraphs are
pentagons and there are no boxes present. Instead we manufactured boxes by 
first applying the $4$-term relation (\ref{4term}). An interesting feature
emerges in (\ref{gpinvf4}). Setting $N$~$=$~$2$ in (\ref{gpinvf4}) the only
non-zero invariants are $T_5$, $T_{71}$ and $T_{94}$ which all evaluate to 
unity. It transpires that for the exceptional group $E_6$, which we consider 
later, these are also the only non-zero invariants present in the 
renormalization group functions. Although their values will not be unity in 
that group. As we will be concentrating on the $N$~$=$~$26$ representation of 
$F_4$ we note that the specific values we used are
\begin{eqnarray}
T_2 &=& 1 ~~,~~
T_3 ~=~ -~ \frac{3}{7} ~~,~~
T_5 ~=~ -~ \frac{25}{98} ~~,~~ 
T_{71} ~=~ \frac{277}{1372} ~~,~~
T_{72} ~=~ -~ \frac{39}{686} \nonumber \\
T_{91} &=& \frac{825}{9604} ~~,~~
T_{92} ~=~ T_{96} ~=~ T_{98} ~=~ -~ \frac{111}{9604} ~~,~~
T_{93} ~=~ T_{95} ~=~ T_{97} ~=~ \frac{39}{9604} \nonumber \\
T_{94} &=& -~ \frac{727}{19208} ~~,~~
T_{99} ~=~ \frac{75}{9604} ~.
\label{gpinvtf426}
\end{eqnarray}
Here we have assumed $T_2$~$=$~$1$ like \cite{25} for the purpose of comparing
our renormalization group functions with that article.  

\sect{$F_4$ renormalization group functions.}

Before examining the consequences of the four loop renormalization group
functions for critical exponent estimates we first record the anomalous
dimensions for the $\phi^2$-type operators in the {\bf 26} and {\bf 324} 
representations of $F_4$. These result from the method outlined in section $2$.
For ${\cal O}^{(26)}_{ij}$ we have\footnote{We have attached an electronic data
file with the operator anomalous dimensions.} 
\begin{eqnarray}
\gamma_{{\cal O}^{(26)}}(g) &=& 
-~ \frac{T_3}{2} g^2 ~+~ \left[ - 7 T_2 T_3 + 18 T_3^2 + 12 T_5 \right] 
\frac{g^4}{48} \nonumber \\ 
&& + \left[ 396 T_2 T_3^2 - 119 T_2^2 T_3 + 198 T_2 T_5 + 486 T_3^3 
+ 2160 \zeta_3 T_3 T_5 - 2556 T_3 T_5
\right. \nonumber \\
&& \left. ~~~ -~ 864 T_{71} - 2592 \zeta_3 T_{72} 
+ 864 T_{72} \right] \frac{g^6}{1728} \nonumber \\
&& + \left[ 1728 \zeta_3 T_2^3 T_3 
- 36961 T_2^3 T_3 
- 15552 \zeta_3 T_2^2 T_3^2 
+ 188556 T_2^2 T_3^2 
+ 48492 T_2^2 T_5
\right. \nonumber \\
&& \left. ~~~ 
+ 23472 T_2 T_3^3
+ 964224 \zeta_3 T_2 T_3 T_5 
- 326592 \zeta_4 T_2 T_3 T_5 
- 820800 T_2 T_3 T_5
\right. \nonumber \\
&& \left. ~~~ 
- 233280 T_2 T_{71}
- 1073088 \zeta_3 T_2 T_{72} 
+ 419904 \zeta_4 T_2 T_{72} 
+ 233280 T_2 T_{72}
\right. \nonumber \\
&& \left. ~~~ 
+ 933120 \zeta_3 T_3^4 
- 1502712 T_3^4
- 4157568 \zeta_3 T_3^2 T_5 
+ 933120 \zeta_4 T_3^2 T_5 
\right. \nonumber \\
&& \left. ~~~ 
+ 3381696 T_3^2 T_5
+ 2737152 \zeta_3 T_3 T_{71} 
- 7464960 \zeta_5 T_3 T_{71} 
+ 3877632 T_3 T_{71}
\right. \nonumber \\
&& \left. ~~~ 
+ 3048192 \zeta_3 T_3 T_{72} 
- 1119744 \zeta_4 T_3 T_{72} 
- 3877632 T_3 T_{72}
- 653184 \zeta_3 T_5^2 
\right. \nonumber \\
&& \left. ~~~ 
+ 660960 T_5^2
+ 3732480 \zeta_3 T_{91} 
- 3732480 \zeta_5 T_{91} 
+ 373248 T_{91}
+ 3732480 \zeta_3 T_{98} 
\right. \nonumber \\
&& \left. ~~~ 
- 3732480 \zeta_5 T_{98} 
- 746496 \zeta_3 T_{99} 
+ 933120 \zeta_5 T_{99} 
+ 1866240 \zeta_3 T_{92} 
\right. \nonumber \\
&& \left. ~~~ 
- 1866240 \zeta_5 T_{92} 
- 4354560 \zeta_3 T_{93} 
+ 6220800 \zeta_5 T_{93} 
+ 1119744 \zeta_3 T_{94} 
\right. \nonumber \\
&& \left. ~~~ 
- 373248 T_{94}
- 6096384 \zeta_3 T_{95} 
+ 8709120 \zeta_5 T_{95} 
+ 3732480 \zeta_3 T_{96} 
\right. \nonumber \\
&& \left. ~~~ 
- 3732480 \zeta_5 T_{96} 
- 2612736 \zeta_3 T_{97} 
+ 3732480 \zeta_5 T_{97} 
\right] \frac{g^8}{746496} ~+~ O( g^{10} ) 
\label{gamma26}
\end{eqnarray}
in the $\MSbar$ scheme where $\zeta(z)$ is the Riemann zeta function. We have 
been able to determine this without reference to specific $F_4$ related group 
identities such as (\ref{4term}) and (\ref{boxdecomp}). In other words to four 
loops the combination of $d^{ijk}$ tensors in each graph could be written in 
terms of one of the group invariants of (\ref{gpinvtdef}). Therefore 
(\ref{gamma26}) can be used for the non-$F_4$ symmetric problems discussed in 
\cite{24}. The situation for the remaining $\phi^2$-type operator is that we 
have had to use $F_4$ based identities as noted earlier. So the four loop 
anomalous dimension is not expressed in terms of the $T_i$ invariants of 
(\ref{gpinvtdef}) and can only be used in the $F_4$ context. In the $\MSbar$
scheme we found 
\begin{eqnarray}
\gamma_{{\cal O}^{(324)}}(g) &=& -~ \frac{g^2}{[N+2]} ~+~ 
[ 5 N + 22 ] \frac{g^4}{24[N+2]^2} \nonumber \\
&& -~ [ 1728 \zeta_3 N^2 - 1465 N^2 - 8640 \zeta_3 N + 13724 N + 7316 ]
\frac{g^6}{864[N+2]^3} \nonumber \\
&& +~ [ 3290976 \zeta_3 N^3 + 641520 \zeta_4 N^3 - 5870880 \zeta_5 N^3
+ 1604411 N^3 - 48224160 \zeta_3 N^2 \nonumber \\
&& ~~~~ - 2426112 \zeta_4 N^2 + 76515840 \zeta_5 N^2 - 17175342 N^2 
+ 105473664 \zeta_3 N \nonumber \\
&& ~~~~ + 1353024 \zeta_4 N - 165939840 \zeta_5 N + 52790148 N 
+ 2595456 \zeta_3 - 3359232 \zeta_4 \nonumber \\
&& ~~~~ - 31726080 \zeta_5 + 53133016 ] \frac{g^8}{373248[N+2]^4} ~+~ 
O(g^{10}) ~.
\label{gamma324}
\end{eqnarray}
The general expressions for $\gamma_\phi(g)$, $\gamma_{\cal O}(g)$ and 
$\beta(g)$ were given in earlier in \cite{24} in terms of the group invariants 
$T_i$ and we do not reproduce them here for $F_4$ as these were given in
\cite{25} at three loops. Instead we have evaluated them for the case when
$N$~$=$~$26$ and together with (\ref{gamma26}) and (\ref{gamma324}) we have 
\begin{eqnarray}
\beta(g) &=& \frac{1}{4} (d-6) g ~-~ \frac{19}{56} g^3 ~-~ 
\frac{1997}{14112} g^5 ~+~ 
[ 3301747 ~-~ 3383856 \zeta_3 ] \frac{g^7}{14224896} \nonumber \\ 
&& +~ [ 1259178385 ~-~ 596452464 \zeta_3 ~+~ 192879792 \zeta_4 ~-~ 
926795520 \zeta_5 ] \frac{g^9}{1194891264} \nonumber \\
&& +~ O(g^{11}) \nonumber \\
\gamma_\phi(g) &=& -~ \frac{1}{12} g^2 ~-~ \frac{149}{3024} g^4 ~-~ 
[ 32400 \zeta_3 ~+~ 78731 ] \frac{g^6}{3048192} \nonumber \\ 
&& +~ [ 3121981 ~-~ 1958160 \zeta_3 ~-~ 256176 \zeta_4 ~-~ 1595520 \zeta_5 ] 
\frac{g^8}{28449792} ~+~ O(g^{10}) \nonumber \\
\gamma_{\cal O}(g) &=& -~ \frac{1}{2} g^2 ~-~ \frac{79}{336} g^4 ~-~ 
[ 16848 \zeta_3 ~+~ 34631 ] \frac{g^6}{84672} \nonumber \\ 
&& +~ [ 230779145 ~-~ 258668208 \zeta_3 ~+~ 21403440 \zeta_4 ~-~ 
89579520 \zeta_5 ] \frac{g^8}{256048128} \nonumber \\
&& +~ O(g^{10}) \nonumber \\
\gamma_{{\cal O}^{(26)}}(g) &=& \frac{3}{14} g^2 ~+~ \frac{53}{784} g^4 ~+~ 
[ 5481 \zeta_3 ~-~ 6689 ] \frac{g^6}{24696} \nonumber \\
&& +~ [ 236544192 \zeta_3 ~-~ 104509440 \zeta_4 ~+~ 413138880 \zeta_5 ~-~ 
531246791 ] \frac{g^8}{597445632} \nonumber \\
&& +~ O(g^{10}) \nonumber \\
\gamma_{{\cal O}^{(324)}}(g) &=& -~ \frac{1}{28} g^2 ~+~ \frac{19}{2352} g^4 
~+~ [ 78275 ~-~ 117936 \zeta_3 ] \frac{g^6}{2370816} \nonumber \\
&& +~ [  218652912 \zeta_3 ~+~ 75524400 \zeta_4 ~-~ 436000320 \zeta_5 ~+~ 
140736511 ] \frac{g^8}{1792336896} \nonumber \\
&& +~ O(g^{10}) 
\end{eqnarray}
using (\ref{gpinvtf426}). Numerically we have 
\begin{eqnarray}
\beta(g) &=& -~ 0.500000 \epsilon g ~-~ 0.339286 g^3 ~-~ 0.141511 g^5 ~-~ 
0.053838 g^6 ~-~ 0.175793 g^8 \nonumber \\
&& +~ O(g^{11}) \nonumber \\
\gamma_\phi(g) &=& -~ 0.083333 g^2 ~-~ 0.049272 g^4 ~-~ 0.038606 g^6 ~-~ 
0.040898 g^8 ~+~ O(g^{10}) \nonumber \\
\gamma_{\cal O}(g) &=& -~ 0.500000 g^2 ~-~ 0.235119 g^4 ~-~ 0.648187 g^6 ~-~ 
0.585346 g^8 ~+~ O(g^{10}) \nonumber \\ 
\gamma_{{\cal O}^{(26)}}(g) &=& 0.214286 g^2 ~+~ 0.067602 g^4 ~-~ 
0.0040705 g^6 ~+~ 0.114445 g^8 ~+~ O(g^{10}) \nonumber \\
\gamma_{{\cal O}^{(324)}}(g) &=& -~ 0.035714 g^2 ~+~ 0.008078 g^4 ~-~ 
0.026780 g^6 ~+~ 0.018529 g^8 ~+~ O(g^{10})
\end{eqnarray}
where the four loop terms are roughly the same magnitude as the lower order 
ones. The three loop values for the first three renormalization group functions
are in agreement with \cite{25}. Clearly there is no sign of a Banks-Zaks fixed
point to four loops with our coupling constant conventions. With the coupling
constant conventions of \cite{25} there are fixed points for the two and four
loop $\beta$-functions but only complex solutions to $\beta(g)$~$=$~$0$ at 
three loops. So there is no robust Banks-Zaks fixed point which would in fact
be an asymptotically safe solution if it had existed. 

\vspace{0.3cm}
{\begin{table}[ht]
\begin{center}
\begin{tabular}{|c||c|c|c|}
\hline
$d$ & $2$ loop & $3$ loop & $4$ loop \\
\hline
$5$ & $1.5731551$ & $1.5613412$ & $1.5639085$ \\ 
$5.95$ & $1.9780535$ & $1.9780512$ & $1.9780512$ \\ 
\hline
\end{tabular}
\end{center}
\begin{center}
{Table $1$. $[0,l]$ Pad\'{e} approximants for $F_4$ exponent $\Delta_\phi$ at 
$l$-loops.}
\end{center}
\end{table}}

Equipped with these renormalization group functions we can evaluate the
$\epsilon$ expansion of the related critical exponents where the fixed point,
$g_c$, is defined by $\beta(g_c)$~$=$~$0$. In order to compare with the results
of \cite{25} we use the notation of that article but define the exponents with
respect to the conventions used here. We recall, \cite{25}, 
\begin{eqnarray}
\Delta_\phi &=& \half d ~-~ 1 ~+~ \gamma_\phi(g_c) \nonumber \\ 
\Delta_{{\cal O}^{(R)}} &=& d ~-~ 2 ~+~ 2 \gamma_\phi(g_c) ~-~ 
2 \gamma_{{\cal O}^{(R)}}(g_c) \nonumber \\
\Delta_{\phi^3} &=& d ~+~ \beta^\prime(g_c) 
\end{eqnarray}
where $\gamma_{{\cal O}^{(1)}}(g_c)$~$\equiv$~$\gamma_{{\cal O}}(g_c)$. Solving
for $g_c$ and evaluating these exponents we find  
\begin{eqnarray}
\Delta_\phi &=& \half d ~-~ 1 ~+~ 0.122807 \epsilon ~-~ 
0.031524 \epsilon^2 ~+~ 0.042483 \epsilon^3 ~+~ 0.122722 \epsilon^4 ~+~ 
O(\epsilon^5) \nonumber \\
\Delta_{{\cal O}^{(1)}} &=& d ~-~ 2 ~-~ 1.228070 \epsilon ~+~ 
0.052388 \epsilon^2 ~-~ 3.414275 \epsilon^3 ~-~ 3.252345 \epsilon^4 ~+~ 
O(\epsilon^5) \nonumber \\
\Delta_{{\cal O}^{(26)}} &=& d ~-~ 2 ~+~ 1.122807 \epsilon ~-~ 
0.031524 \epsilon^2 ~+~ 0.042483 \epsilon^3 ~+~ 0.122722 \epsilon^4 ~+~
O(\epsilon^5) \nonumber \\
\Delta_{{\cal O}^{(324)}} &=& d ~-~ 2 ~+~ 0.140351 \epsilon ~-~ 
0.162835 \epsilon^2 ~-~ 0.172846 \epsilon^3 ~-~ 0.472810 \epsilon^4 ~+~
O(\epsilon^5) \nonumber \\
\Delta_{\phi^3} &=& d ~+~ 2 \epsilon ~-~ 1.229301 \epsilon^2 ~-~ 
0.132727 \epsilon^3 ~-~ 8.882515 \epsilon^4 ~+~ O(\epsilon^5)
\end{eqnarray} 
where the terms to three loops of $\Delta_\phi$, $\Delta_{{\cal O}^{(1)}}$ and 
$\Delta_{\phi^3}$ are in agreement with \cite{25}. Also the coefficients of
$\Delta_\phi$ and $\Delta_{{\cal O}^{(26)}}$ are consistent with 
${\cal O}^{(26)}$ being a conformal descendant of $\phi^i$. Surprisingly the 
four loop correction to $\Delta_{\phi^3}$ is large. One of the main features of
\cite{25} was the comparison of the exponent $\Delta_\phi$ with the value 
obtained from the conformal bootstrap analysis for dimensions $d$~$=$~$5$ and 
$5.95$. In Tables $1$ and $2$ we have provided estimates for $\Delta_\phi$ 
using Pad\'{e} approximants at successive loop orders in order to gauge 
convergence. In Table $1$ we note the $[0,l]$ estimate at the $l$-loop order. 
The results for $d$~$=$~$5.95$ dimension appear to converge while those for 
$d$~$=$~$5$ have not converged as well but do appear to have settled to a value
in the neighbourhood of $1.56$. In compiling the Pad\'{e} analysis what was 
apparent was that the estimates from the other $[p,q]$ approximants were not 
dissimilar to the $[0,l]$ ones. Therefore to improve estimates we calculated 
the average of all the approximants at each loop. These are presented in Table 
$2$ and indicates a four loop value closer to $1.55$.

\vspace{0.3cm}
{\begin{table}[ht]
\begin{center}
\begin{tabular}{|c||c|c|c|}
\hline
$d$ & $2$ loop & $3$ loop & $4$ loop \\
\hline
$5$ & $1.5632667$ & $1.5589076$ & $1.5516367$ \\ 
$5.95$ & $1.9780520$ & $1.9780511$ & $1.9780512$ \\ 
\hline
\end{tabular}
\end{center}
\begin{center}
{Table $2$. Average of Pad\'{e} approximants at each loop order for $F_4$ 
exponent $\Delta_\phi$.}
\end{center}
\end{table}}

\vspace{0.3cm}
{\begin{table}[ht]
\begin{center}
\begin{tabular}{|c|c||c|c|c|}
\hline
$d$ & $R$& $2$ loop & $3$ loop & $4$ loop \\
\hline
$5$ & $\mathbf{26}$ & $3.5598518$ & $3.5585128$ & $3.5648212$ \\ 
    & $\mathbf{324}$ & $3.0848268$ & $3.0317658$ & $2.9986463$ \\ 
\hline
$5.95$ & $\mathbf{26}$ & $3.9780513$ & $3.9780511$ & $3.9780512$ \\ 
       & $\mathbf{324}$ & $3.9534156$ & $3.9534045$ & $3.9534041$ \\ 
\hline
\end{tabular}
\end{center}
\begin{center}
{Table $3$. $[0,l]$ Pad\'{e} approximants for $F_4$ exponent
$\Delta_{{\cal O}^{(R)}}$ at $l$-loops in dimension $d$.}
\end{center}
\end{table}}

Given these estimates for $\Delta_\phi$ we have repeated the same Pad\'{e} 
analysis for the $\phi^2$-type operators in the {\bf 26} and {\bf 324} 
representations as these are the only cases with non-zero critical exponents. 
The results are presented in Tables $3$ and $4$ where we gather the $5$ and 
$5.95$ dimension estimates in each table. The former Table has the values for 
the $[0,l]$ Pad\'{e} approximants at $l$-loops and the latter has the average 
of the Pad\'{e}'s at each loop order. For $5.95$ dimensions the exponents agree
to at least four decimal places for each of the representations. In five 
dimensions the convergence is not as fast but again there appears to be a 
consistent value to two decimal places. In terms of comparing the exponent 
values in different representations in a particular dimension the operators are
virtually degenerate in $5.95$ dimensions. The discrepancy between them is 
around $0.5\%$. As the spacetime dimension decreases this effective degeneracy 
is lifted. It transpires that the critical exponent for the {\bf 324} 
representation, $\Delta_{{\cal O}^{(324)}}$, is lower than that of the {\bf 26}
representation, $\Delta_{{\cal O}^{(26)}}$. Moreover the former decreases more 
rapidly than the one for the {\bf 26} representation as the spacetime dimension
decreases. A similar feature was evident in the analysis of \cite{25}.

It is now instructive to compare our four loop estimates with the conformal 
bootstrap results of \cite{25}. In that article plots were given of the allowed
and excluded regions of the parameter space defined by $\Delta_\phi$ ($x$-axis)
and what was termed $\Delta^{2\mbox{\footnotesize{nd}}}_{\bf 26}$ ($y$-axis) in
the notation of \cite{25}. In the bootstrap approach the location of a kink in 
the boundary of these two regions is the point where one can read off an 
estimate for $\Delta_\phi$ for example from the $x$-coordinate. First, the plot
of \cite{25} in $5.95$ dimensions, for instance, has a well defined kink with 
an $x$-coordinate value of $1.978$. This is in accord with the three loop 
$\epsilon$ series estimates provided in \cite{25}. Our new four loop values in 
Tables $1$ and $2$ are not inconsistent with this. In addition what is apparent
from the plot of \cite{25} in this spacetime dimension is that the 
$y$-coordinate corresponds to a value fractionally shy of $4$. This is the 
estimate given in \cite{25} for the quantity
$\Delta^{2\mbox{\footnotesize{nd}}}_{\bf 26}$. This is not dissimilar to the 
estimates of both $\Delta_{{\cal O}^{(26)}}$ and 
$\Delta_{{\cal O}^{(324)}}$ shown in Tables $3$ and $4$. Indeed our four loop
estimates show a small change from the one loop estimates given in Table $1$ of
\cite{25} for these exponents.

The situation for the conformal bootstrap analysis in five dimensions is 
different. While there is an allowed and excluded region in the corresponding
plot of \cite{25} there are no sharp kinks or boundaries. Instead there are
what was termed weak kinks, \cite{25}. From Figure $4$ of \cite{25} the kinks 
have slope changes at about $1.55$ and $1.6$ in the $x$-coordinate. Indeed the 
latter value is what was quoted for the bootstrap estimate for $\Delta_\phi$. 
However, from Table $2$ the average four loop Pad\'{e} estimate for 
$\Delta_\phi$ falls closer to the value of $1.55$ rather than the quoted value 
of $1.6$ in \cite{25}. Indeed it was noted in \cite{25} that this latter value 
was not in full agreement with the three loop perturbation theory used in 
\cite{25}. Our four loop result shows that the series for $\Delta_\phi$ is not 
diverging as is apparent from Table $2$. Put another way if one regards the 
{\em lower} kink of Figure $4$ of \cite{25} as the one to be used for 
estimating exponents then the perturbative result is not inconsistent with the 
bootstrap technology. To support this point of view one can examine the 
location of the first weak kink or lower knuckle of Figure $4$ in \cite{25} 
with respect to the $y$-axis. This is roughly at $3.1$, \cite{25}. Our 
estimates for $\Delta_{{\cal O}^{(324)}}$ are in the region of $3.01$ which 
appears to be consistent with the $y$-axis value of Figure $4$ in \cite{25}. 
For the {\bf 26} representation we find an estimate for the 
$\Delta_{{\cal O}^{(26)}}$ exponent of around $3.55$. This is lower than the 
upper knuckle of the five dimensional plot in \cite{25} which appears closer to
$3.9$. However, the three curves presented in Figure $4$ of \cite{25} have not 
converged to the same accuracy as those in the neighbourhood of the lower 
knuckle in $5$ dimensions or indeed that for $5.95$ dimensions. Moreover in the 
latter spacetime dimension it is the lower corner of the plot of Figure $4$ of 
\cite{25} which gives the dimension of either $\phi^2$-type operator. That
should also be the case in $5$ dimensions in order to have a consistent point 
of view. What is interesting is that the conformal bootstrap analysis appears 
to give relatively accurate data on the exponent of the operator with the 
lowest value. However if one wishes to marry the information derived from 
perturbation theory here with the data from the $5$ dimensional bootstrap 
analysis then one would have to regard the exponent estimate from the lower 
kink as corresponding to that of the $\phi^2$-type operator in the {\bf 324} 
representation. 

\vspace{0.3cm}
{\begin{table}[ht]
\begin{center}
\begin{tabular}{|c|c||c|c|c|}
\hline
$d$ & $R$& $2$ loop & $3$ loop & $4$ loop \\
\hline
$5$ & $\mathbf{26}$ & $3.5566150$ & $3.5578930$ & $3.5509073$ \\ 
    & $\mathbf{324}$ & $3.0562147$ & $3.0062392$ & $3.0267278$ \\ 
\hline
$5.95$ & $\mathbf{26}$ & $3.9780509$ & $3.9780511$ & $3.9780512$ \\ 
       & $\mathbf{324}$ & $3.9534112$ & $3.9534043$ & $3.9534041$ \\ 
\hline
\end{tabular}
\end{center}
\begin{center}
{Table $4$. Average of Pad\'{e} approximants at each loop order for $F_4$ 
exponent $\Delta_{{\cal O}^{(R)}}$ at $l$-loops in dimension $d$.}
\end{center}
\end{table}}

\sect{$E_6$.}

For the remaining part of our study of $\phi^3$ theory with exceptional
symmetry we concentrate on the group $E_6$ where the fundamental representation
is {\bf 27} and the adjoint is {\bf 78}. As $E_6$ is a complex group then the
Lagrangian for a cubic theory with $E_6$ symmetry involves fields $\phi^i$ and 
$\bar{\phi}^i$ and the tensors $d_{ijk}$ and $d^{ijk}$. We take the convention 
that the conjugate to $d^{ijk}$ is $d_{ijk}$ similar to \cite{25}. Therefore 
the $E_6$ symmetric Lagrangian is 
\begin{equation}
L ~=~ \partial_\mu \bar{\phi}_i \partial^\mu \phi^i ~+~
\frac{g}{6} \left( d_{ijk} \phi^i \phi^j \phi^k ~+~
d^{ijk} \bar{\phi}_i \bar{\phi}_j \bar{\phi}_k \right) ~.
\label{lage6}
\end{equation}
This is similar in structure to the cubic theory with $SU(3)$~$\times$~$SU(3)$
symmetry considered in \cite{37,38}. Moreover, the Feynman graphs generated
from (\ref{lage6}) share properties similar to those of the
$SU(3)$~$\times$~$SU(3)$ theory. The main one is that there are no Feynman
diagrams with subgraphs with an odd number of legs. So for instance there is
no one loop triangle graph for the renormalization of the coupling constant.
It is straightforward to establish this by realising that the two vertices
of (\ref{lage6}) are what is termed directed. Either all the arrows indicating 
the charge flow on each vertex line points to the interaction point itself or
points away. Thus it is easy to see that in a one loop triangle graph the lines
cannot be decorated with arrows which point to or from all the vertices. To
reflect this aspect of the properties of the $E_6$ Lagrangian the indices of
the tensor associated with the coupling constant are either raised or lowered,
\cite{25}. This convention will only be applied in this section. Moreover we 
will use upper and lower group indices on the fields themselves in keeping with
the notion of distinguishing that there is a flow of charge in contrast to 
$F_4$.

To construct the four loop $E_6$ renormalization group functions we need to 
determine the values of the group invariants. The properties of the $E_6$ Lie 
algebra differ from those of $F_4$ but we will use the same algorithm as before 
to derive an identity for the one loop box akin to (\ref{boxdecomp}). We base 
our derivations on $E_6$ group properties derived in \cite{39} which used the 
more mathematical articles \cite{40,41,42,43}. Further background to the 
structure and properties of $E_6$ can be found in \cite{30,44}. In \cite{39} 
the convention for the product of tensors was 
\begin{equation}
d_{ikl} d^{jkl} ~=~ 10 \delta_i^{~j}
\end{equation}
which imples $T_2$~$=$~$10$. From this and identities derived in 
\cite{40,41,42,43} it was shown in \cite{39} that 
\begin{equation}
d_{i i_1 i_3} d^{j i_2 i_3} d_{k i_2 i_4} d^{l i_1 i_4}
~=~ 5 \left[ \delta_i^{~ j} \delta_k^{~ l} ~+~
\delta_i^{~ l} \delta_k^{~ j} \right] ~-~ 4 d_{i k i_1} d^{j l i_1} ~.
\end{equation}
This is the $E_6$ equivalent of the one loop box topology of (\ref{boxdecomp}).
Using this we have determined the values of all the invariants of 
(\ref{gpinvtdef}). Before recording the values we note that for topologies 
where at least one of the one loop subgraphs has an odd number of external legs
it is not possible to decorate the lines consistently in such a way that all
the vertices have all arrows pointing in or out. In these cases the invariant
is set to zero as such Feynman graphs would not be generated in the first
place using, say, the {\sc Qgraf} package, \cite{31}. It transpires that to
four loops there are only three non-zero invariants aside from $T_2$. In
summary the values are  
\begin{equation}
T_2 ~=~ 10 ~~~,~~~ 
T_5 ~=~ -~ 30 ~~~,~~~ 
T_{71} ~=~ 220 ~~~,~~~ 
T_{94} ~=~ -~ 530
\end{equation}
after applying the box rule to (\ref{gpinvtdef}). 

It is a straightforward exercise to substitute these values in the expressions
for the renormalization group functions which have been expressed in terms of
$T_i$ to find 
\begin{eqnarray}
\beta(g) &=& \frac{1}{4} (d-6) g ~-~ \frac{5}{4} g^3 ~+~ \frac{265}{72} g^5 ~+~ 
5 [ 163183 ~-~ 19440 \zeta_3 ] \frac{g^7}{5184} \nonumber \\
&& +~ 5 [ 1044144 \zeta_3 ~+~ 32400 \zeta_4 ~-~ 633600 \zeta_5 ~+~ 591527 ] 
\frac{g^9}{3456} ~+~ O(g^{11}) \nonumber \\
\gamma_\phi(g) &=& -~ \frac{5}{6} g^2 ~-~ \frac{275}{108} g^4 ~+~ 
25 [ 2699 ~-~ 3888 \zeta_3 ] \frac{g^6}{7776} \nonumber \\
&& +~ 25 [ 143119 ~-~ 9936 \zeta_3 ~+~ 6480 \zeta_4 ~-~ 126720 \zeta_5 ]
\frac{g^8}{5184} ~+~ O(g^{10}) \nonumber \\
\gamma_{\cal O}(g) &=& -~ 5 g^2 ~-~ \frac{25}{12} g^4 ~-~ \frac{9575}{108} g^6 
\nonumber \\
&& +~ 25 [ 3401136 \zeta_3 ~-~ 174960 \zeta_4 ~-~ 10264320 \zeta_5 ~+~ 
13869707 ] \frac{g^8}{46656} ~+~ O(g^{10}) \nonumber \\ 
\gamma_{{\cal O}^{(27)}}(g) &=& -~ \frac{15}{2} g^4 ~-~ 
\frac{1155}{8} g^6 ~+~ 1055 [ 49 ~-~ 432 \zeta_3 ] \frac{g^8}{288} ~+~ 
O(g^{10})
\end{eqnarray}
where there is no one loop contribution to the final renormalization group
function. Numerically we have 
\begin{eqnarray}
\beta(g) &=& -~ 0.500000 \epsilon ~-~ 1.250000 g^3 ~+~ 3.680556 g^5 ~+~ 
134.852444 g^7 ~+~ 1771.871166 g^9 \nonumber \\
&& +~ O(g^{11}) \nonumber \\
\gamma_\phi(g) &=& -~ 0.833333 g^2 ~-~ 2.546296 g^4 ~-~ 6.348371 g^6 ~+~
32.741763 g^8 ~+~ O(g^{10}) \nonumber \\
\gamma_{\cal O}(g) &=& -~ 5.000000 g^2 ~-~ 2.083333 g^4 ~-~ 88.657407 g^6 ~+~ 
3818.021497 g^8 ~+~ O(g^{10}) \nonumber \\
\gamma_{{\cal O}^{(27)}}(g) &=& -~ 7.500000 g^4 ~-~ 144.375000 g^6 ~-~ 
1722.758521 g^8 ~+~ O(g^{10}) ~. 
\end{eqnarray}
In \cite{25} it was noted that from the one loop $\beta$-function there was a
stable unitary fixed point which we confirm here allowing for the different
convention on the definition of the sign of our coupling constant in 
(\ref{lag}). Compared to the $F_4$ $\beta$-function the coefficients of the
$E_6$ $\beta$-function increase significantly with the loop order. This can be
traced, however, to the different values of $T_2$ which is $10$ for $E_6$ 
instead of unity for $F_4$. If one rescaled $g^2$ by a factor of $10$ then the
coefficients of $\beta(g)$ would be comparable to those of $F_4$. In our 
coupling constant conventions the $E_6$ case like $F_4$ exhibits asymptotic 
freedom and to four loops has a Banks-Zaks fixed point. At two, three and four 
loops this is at $g^2$~$=$~$0.339623$, $0.083593$ and $0.063944$ respectively. 
The latter values suggest convergence. At three loops there is a fixed point 
for negative coupling which is not present at two or four loops. If it had been
present in those cases then $E_6$ could be a model with the property of 
asymptotic safety. 

In advance of a conformal bootstrap analysis we can now provide the $\epsilon$
expansion of the related critical exponents at the Wilson-Fisher fixed point at
four loops. These are 
\begin{eqnarray}
\Delta_\phi &=& \half d ~-~ 1 ~+~ 0.333333 \epsilon ~-~ 0.800000 \epsilon^2 
~+~ 8.044444 \epsilon^3 ~-~ 84.333501 \epsilon^4 ~+~ O(\epsilon^5) \nonumber \\
\Delta_{{\cal O}^{(1)}} &=& d ~-~ 2 ~-~ 3.333333 \epsilon ~+~ 
3.777778 \epsilon^2 ~-~ 76.971365 \epsilon^3 + 505.735425 \epsilon^4 ~+~ 
O(\epsilon^5) \nonumber \\
\Delta_{{\cal O}^{(27)}} &=& d ~-~ 2 ~+~ 1.333333 \epsilon ~-~ 
0.800000 \epsilon^2 ~+~ 8.044444 \epsilon^3 ~-~ 84.333501 \epsilon^4 ~+~
O(\epsilon^5) \nonumber \\
\Delta_{\phi^3} &=& d ~+~ 2 \epsilon ~+~ 2.355556 \epsilon^2 ~-~ 
74.593093 \epsilon^3 ~+~ 885.932572 \epsilon^4 ~+~ O(\epsilon^5) ~.
\end{eqnarray}
It transpires that the respective coefficients of the exponents are much larger
than their $F_4$ counterparts. In effect what this means is that estimates for 
$E_6$ exponents from perturbation theory may only be reliable for a value of 
$d$ relatively close to six.

\sect{Discussion.}

Our original aim was to extend the three loop analysis of $F_4$ symmetric
scalar cubic theory in six dimensions to four loops. Having achieved this we
derived critical exponents for the field and $\phi^2$-type operators in various
representations of $F_4$ to the same order of precision. This is important in
the context of predictions from the conformal bootstrap analysis of \cite{25}.
In that paper there was a suggestion that the difference in the $d$~$=$~$5$
dimension estimate for $\Delta_\phi$ compared to perturbative prediction was 
due to non-perturbative effects. From the Pad\'{e} analysis we noted that the 
exponent derived from the $O(\epsilon^4)$ correction was smaller than the three
loop result and, moreover, closer to that for the bootstrap. Although in 
$d$~$=$~$5$ there was not as sharp an estimate compared to the $d$~$=$~$5.95$ 
dimensional case. From the perturbative side the estimates for the dimension 
two operator exponents determined the order of their appearance in the 
spectrum. Close to six dimensions the two operators were effectively degenerate
and the perturbative estimate for their exponents was in sharp agreement with 
\cite{25}. For the lower dimensional case studied in \cite{25} the value
recorded there for $\Delta^{2\mbox{\footnotesize{nd}}}_{\bf 26}$ was consistent
with the estimate for the $\phi^2$-type operator in the {\bf 324} 
representation rather than that in the {\bf 26} representation. It would be 
interesting to have a conformal bootstrap analysis for the exceptional group 
$E_6$. The observation of \cite{25} that there appears to be a stable infrared 
fixed point in five dimensions seems to be confirmed. However, such a bootstrap
analysis could give further insight into the role of the analogous 
$\phi^2$-type operator in $E_6$. From properties of the $E_6$ group it would 
appear that there is no operator parallel to that in the {\bf 324} 
representation of $F_4$. Instead there is only the {\bf 27} one. In other words
a conformal bootstrap analysis should be able to estimate the $E_6$ value for 
what would be the exponent $\Delta^{2\mbox{\footnotesize{nd}}}_{\bf 27}$ 
accurately and then compare with the four loop $\epsilon$ expansion. Equally
the other groups in the $F_4$ family can be analysed by the bootstrap given 
that perturbative results are now available to the same accuracy as the $F_4$
case itself.

\vspace{1cm}
\noindent
{\bf Acknowledgements.} The author thanks R.M. Simms for discussions. The work 
was carried out with the support of the STFC through the Consolidated Grant 
ST/L000431/1.

\appendix 

\sect{Exponents for related $F_4$ groups.}

While we have focused in the main on the exceptional group $F_4$ there are
several other groups which have a tensor $d^{ijk}$ which satisfies the 
$4$-term relation of (\ref{4term}) and
\begin{equation}
d^{ijj} ~=~ 0 ~~~~,~~~~ d^{ikl} d^{jkl} ~=~ T_2 \delta^{ij} ~.
\end{equation}  
Using the Lie algebra classification notation these are $A_1$, $A_2$ and $C_3$,
\cite{25,30}, where $A_1$ is also equivalent to $B_1$. Values of the critical
exponents for $\phi^3$ theory based on these symmetry groups can be deduced
from the results of section $4$ by setting $N$~$=$~$5$, $8$ and $14$ 
respectively. Thus we have extended the three loop results of \cite{25} for
each case and found  
\begin{eqnarray}
\Delta_\phi &=& \half d ~-~ 1 ~+~ 0.179487 \epsilon ~+~ 0.174885 \epsilon^2 ~+~
1.446636 \epsilon^3 ~+~ 11.125264 \epsilon^4 ~+~ O(\epsilon^5) \nonumber \\
\Delta_{{\cal O}^{(1)}} &=& d ~-~ 2 ~-~ 1.794872 \epsilon ~-~ 
2.641683 \epsilon^2 ~-~ 25.875476 \epsilon^3 ~-~ 179.737315 \epsilon^4 ~+~ 
O(\epsilon^5) \nonumber \\
\Delta_{{\cal O}^{(5)}} &=& d ~-~ 2 ~+~ 1.179487 \epsilon ~+~ 
0.174885 \epsilon^2 ~+~ 1.446636 \epsilon^3 ~+~ 11.125264 \epsilon^4 ~+~ 
O(\epsilon^5) \nonumber \\
\Delta_{{\cal O}^{(9)}} &=& d ~-~ 2 ~-~ 0.256410 \epsilon ~-~ 
1.214990 \epsilon^2 ~-~ 7.225638 \epsilon^3 ~-~ 54.804917 \epsilon^4 ~+~ 
O(\epsilon^5) \nonumber \\
\Delta_{\phi^3} &=& d ~+~ 2 \epsilon ~-~ 3.880342 \epsilon^2 ~-~ 
24.132914 \epsilon^3 ~-~ 250.285961 \epsilon^4 ~+~ O(\epsilon^5)
\end{eqnarray} 
for $N$~$=$~$5$ corresponding to $A_1$. For $N$~$=$~$8$ which relates to the 
Lie algebra $A_2$ we have 
\begin{eqnarray}
\Delta_\phi &=& \half d ~-~ 1 ~+~ 0.151515 \epsilon ~+~ 0.041740 \epsilon^2 ~+~
0.397533 \epsilon^3 ~+~ 2.024208 \epsilon^4 ~+~ O(\epsilon^5) \nonumber \\
\Delta_{{\cal O}^{(1)}} &=& d ~-~ 2 ~-~ 1.515152 \epsilon ~-~ 
0.959179 \epsilon^2 ~-~ 10.049808 \epsilon^3 ~-~ 38.922333 \epsilon^4 ~+~ 
O(\epsilon^5) \nonumber \\
\Delta_{{\cal O}^{(8)}} &=& d ~-~ 2 ~+~ 1.151515 \epsilon ~+~ 
0.041740 \epsilon^2 ~+~ 0.397533 \epsilon^3 ~+~ 2.024208 \epsilon^4 ~+~ 
O(\epsilon^5) \nonumber \\
\Delta_{{\cal O}^{(27)}} &=& d ~-~ 2 ~-~ 0.060606 \epsilon ~-~ 
0.521746 \epsilon^2 ~-~ 1.853121 \epsilon^3 ~-~ 9.180040 \epsilon^4 ~+~ 
O(\epsilon^5) \nonumber \\
\Delta_{\phi^3} &=& d ~+~ 2 \epsilon ~-~ 2.389348 \epsilon^2 ~-~ 
7.729109 \epsilon^3 ~-~ 65.712863 \epsilon^4 ~+~ O(\epsilon^5) ~. 
\end{eqnarray} 
We note that the sign of the $O(\epsilon^3)$ term of $\Delta_{{\cal O}^{(1)}}$ 
differs from that given in \cite{25} which we assume is a typographical error.
Finally, we find 
\begin{eqnarray}
\Delta_\phi &=& \half d ~-~ 1 ~+~ 0.133333 \epsilon ~-~ 0.011111 \epsilon^2 ~+~
0.120049 \epsilon^3 ~+~ 0.437738 \epsilon^4 ~+~ O(\epsilon^5) \nonumber \\
\Delta_{{\cal O}^{(1)}} &=& d ~-~ 2 ~-~ 1.333333 \epsilon ~-~ 
0.244444 \epsilon^2 ~-~ 5.098686 \epsilon^3 ~-~ 10.110099 \epsilon^4 ~+~ 
O(\epsilon^5) \nonumber \\
\Delta_{{\cal O}^{(14)}} &=& d ~-~ 2 ~+~ 1.133333 \epsilon ~-~ 
0.011111 \epsilon^2 ~+~ 0.120049 \epsilon^3 ~+~ 0.437738 \epsilon^4 ~+~ 
O(\epsilon^5) \nonumber \\
\Delta_{{\cal O}^{(90)}} &=& d ~-~ 2 ~+~ 0.066667 \epsilon ~-~ 
0.260000 \epsilon^2 ~-~ 0.520808 \epsilon^3 ~-~ 1.816381 \epsilon^4 ~+~ 
O(\epsilon^5) \nonumber \\
\Delta_{\phi^3} &=& d ~+~ 2 \epsilon ~-~ 1.611111 \epsilon^2 ~-~ 
2.094961 \epsilon^3 ~-~ 20.962996 \epsilon^4 ~+~ O(\epsilon^5)
\end{eqnarray} 
for $C_3$ having set $N$~$=$~$14$. The dimensions of ${\cal O}^{(R)}$ for 
$R$~$\neq$~$1$ are determined from the respective $F_4$ expressions for the 
{\bf 26} and {\bf 324} representations which were computed as functions of $N$.
Our expressions for $\Delta_\phi$, $\Delta_{{\cal O}^{(1)}}$ and 
$\Delta_{\phi^3}$ agree with the three loop ones given in the Appendix of 
\cite{25}, aside from the one noted above, but the four loop contributions are 
new. Also within our conventions the coefficients in the $\epsilon$ expansion 
of the $\phi^2$-type operators derived from the {\bf 26} representation of 
$F_4$ are in accord with $\Delta_\phi$ consistent with the relation to the 
conformal descendant operator.

\end{document}